\documentclass[manuscript]{aastex}
\usepackage{graphicx,natbib,color}
\usepackage[T1]{fontenc} 
\shorttitle{Double-Streamer/Pseudostreamer Hybrid}
\shortauthors{Rachmeler et al.}

\begin{document}

\title{Observations of a Hybrid Double-Streamer/Pseudostreamer in the
  Solar Corona}


\author{L.~A. Rachmeler\altaffilmark{1},
  S.~J. Platten\altaffilmark{2}, C. Bethge\altaffilmark{3},
  D.~B. Seaton\altaffilmark{1}, and A.~R. Yeates\altaffilmark{4}}
\affil{$^{1}$Royal Observatory of Belgium, Avenue Circulaire 3, 1180 Brussels, Belgium}
\affil{$^{2}$School of Mathematics and Statistics, University of St. Andrews, North Haugh, St. Andrews, Fife KY16 9SS, UK }
\affil{$^{3}$Kiepenheuer-Institut f\"{u}r Sonnenphysik, Sch\"{o}neckstr. 6, 79104 Freiburg, Germany}
\affil{$^{4}$Department of Mathematical Sciences, Durham University, Science Laboratories, South Road, Durham DH1 3LE, UK}
\email{rachmeler@oma.be}


\begin{abstract}
  We report on the first observation of a single hybrid magnetic
  structure that contains both a pseudostreamer and a double helmet
  streamer. This structure was originally observed by the SWAP
  instrument aboard the PROBA2 satellite between 5 and 10~May~2013. It
  consists of a pair of filament channels near the south pole of the
  sun. On the western edge of the structure, the magnetic morphology
  above the filaments is that of a side-by-side double helmet
  streamer, with open field between the two channels. On the eastern
  edge, the magnetic morphology is that of a coronal pseudostreamer
  without the central open field. We investigated this structure with
  multiple observations and modelling techniques. We describe the
  topology and dynamic consequences of such a unified structure.
\end{abstract}



\section{Introduction} \label{sec:intro}

Streamer-like structures have been studied for many years, and there
are two general categories of features that fall into this
classification: streamers and pseudostreamers/unipolar streamers
\citep{pneuman1971, wang2007}. They are often identified by their
upper-coronal white-light signatures, which are both extended bright
radial features. Although these radial patterns are slightly different, it
is the magnetic morphologies that truly distinguish the two types of
structures. Those morphologies cannot be pinpointed with white-light
measurements, partially due to the spatial gap in data coverage, but
also because the measurements are sensitive to density, rather than
magnetic field. Similarly, coronal EUV imagers record properties of
the plasma, which follows the magnetic field, but are not directly
sensitive to it.

In this letter, we focus specifically on the magnetic properties of
the two different morphologies. For clarity, and because the
definitions vary slightly in the current literature, we
define the two structures as follows: A coronal \emph{streamer} is a
magnetic structure overlying a single (or an odd number of) polarity
inversion lines (PILs) with closed loops in the lower corona and
oppositely oriented open magnetic field in the upper corona, such that
a current sheet and plasma sheet are present between the two open
field domains. A \emph{pseudostreamer} is a magnetic structure
overlying two (or an even number of) PILs such that above the
closed field, two domains of open field of the same polarity come
together and no current sheet is present.

We present the first identification of a single hybrid
magnetic structure composed of both a side-by-side double streamer
(DS) and a pseudostreamer (PS). The transition between the two
morphologies occurs in space, rather than in time, such that both
exist simultaneously within a coherent structure that changes along
its length. The change of magnetic topology has implications for the
stability of the enclosed filament channels, and the solar wind
properties of the system.

Pseudostreamers can be very long lived. We observed a PS in the
southern hemisphere of the sun that persisted from at least mid-2012
until mid-2013. This structure can be seen in PROBA2/SWAP movies,
rotating in and out of the plane-of-sky (see Figure 5 
by \citealp{seaton2013.2}). The PS is generally only visible for part
of the rotation, although one or both of the individual filament
channels can extend beyond the region where the PS topology is
present. In the case where both filament channels extend beyond the
PS, it is possible for the PS to split into a DS.

We have observed an example of just this type of topological
change. Near the leading edge of the long-lived PS discussed above, we
observed a DS that persisted for several rotations. In this letter we
present these observations during Carrington rotation~2136 in
May~2013. We also present simulated observations of simple magnetic
models that support our interpretation of this structure.

\section{Magnetic models and morphology} \label{sec:topology}

\begin{figure}

   \centerline{\hspace*{0.015\textwidth}
     \hspace*{+0.001\textwidth}\includegraphics[width=0.3\textwidth,clip=]{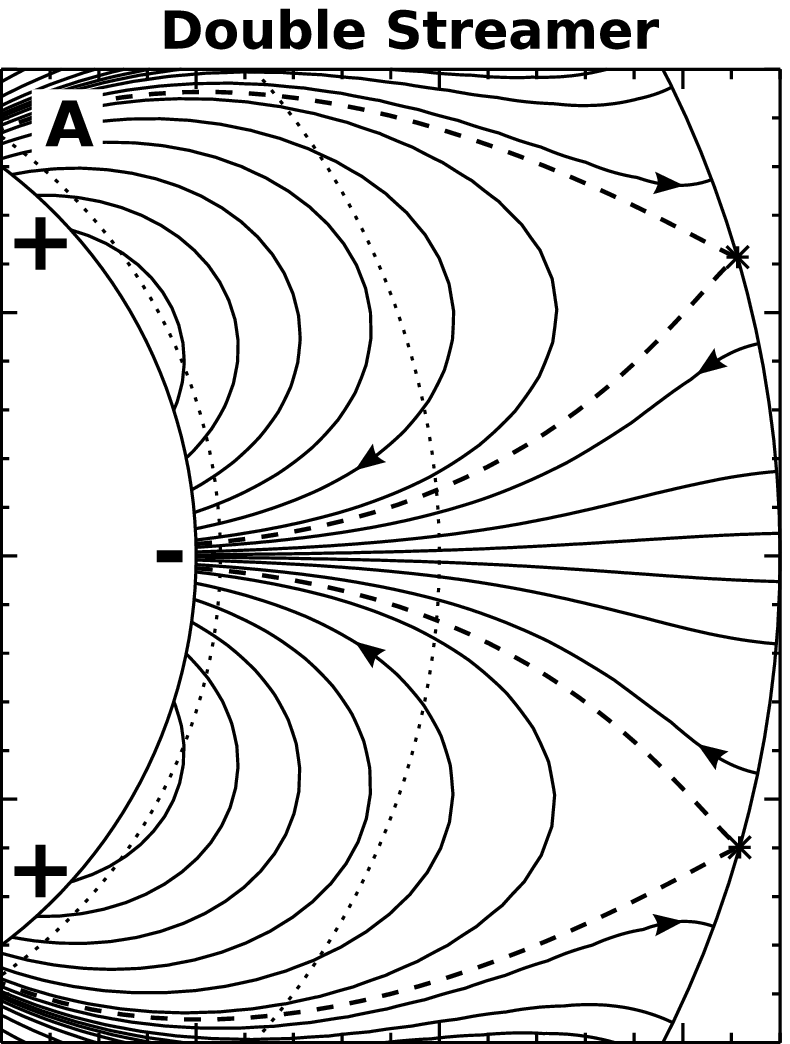}
               \hspace*{+0.08\textwidth}
               \includegraphics[width=0.3\textwidth,clip=]{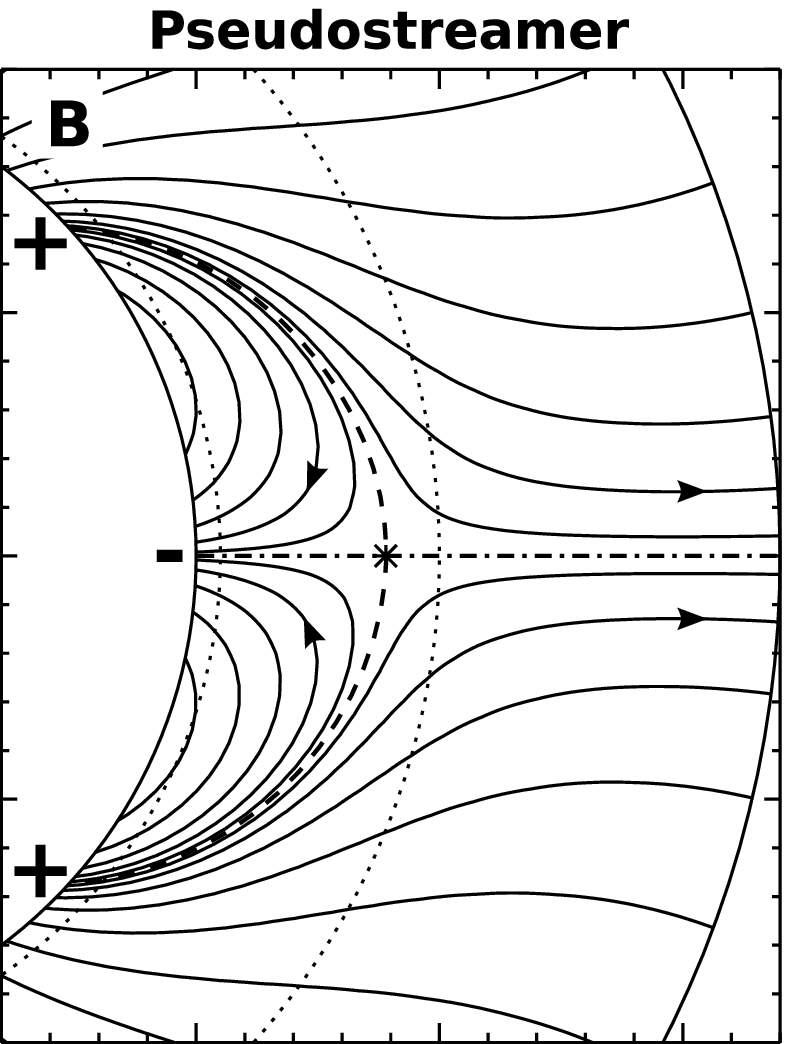}
              }
              \vspace{-0.05\textwidth}
   \centerline{\hspace*{0.015\textwidth}
     \includegraphics[width=0.53\textwidth,clip=]{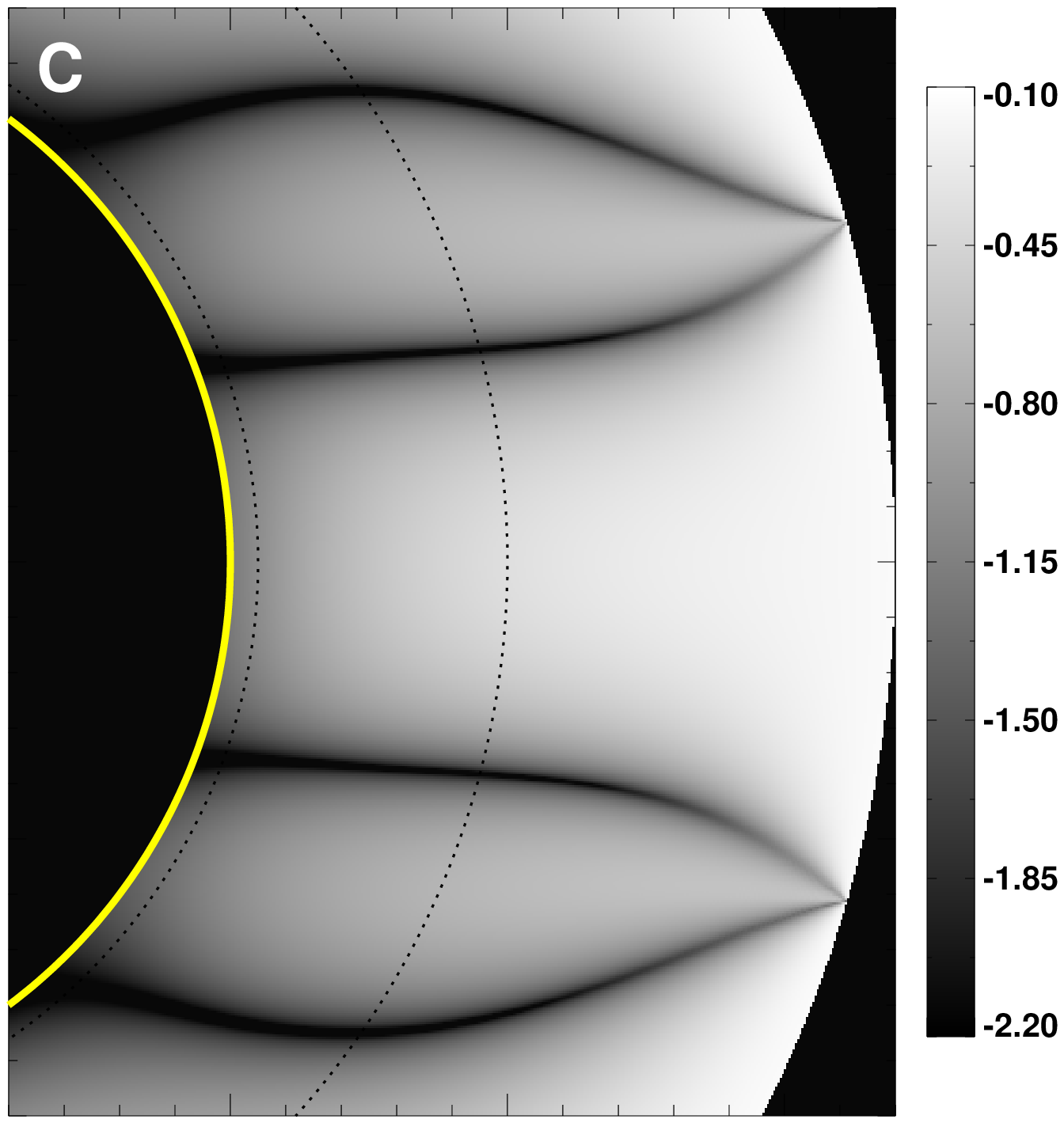}
               \hspace*{-0.15\textwidth}
               \includegraphics[width=0.53\textwidth,clip=]{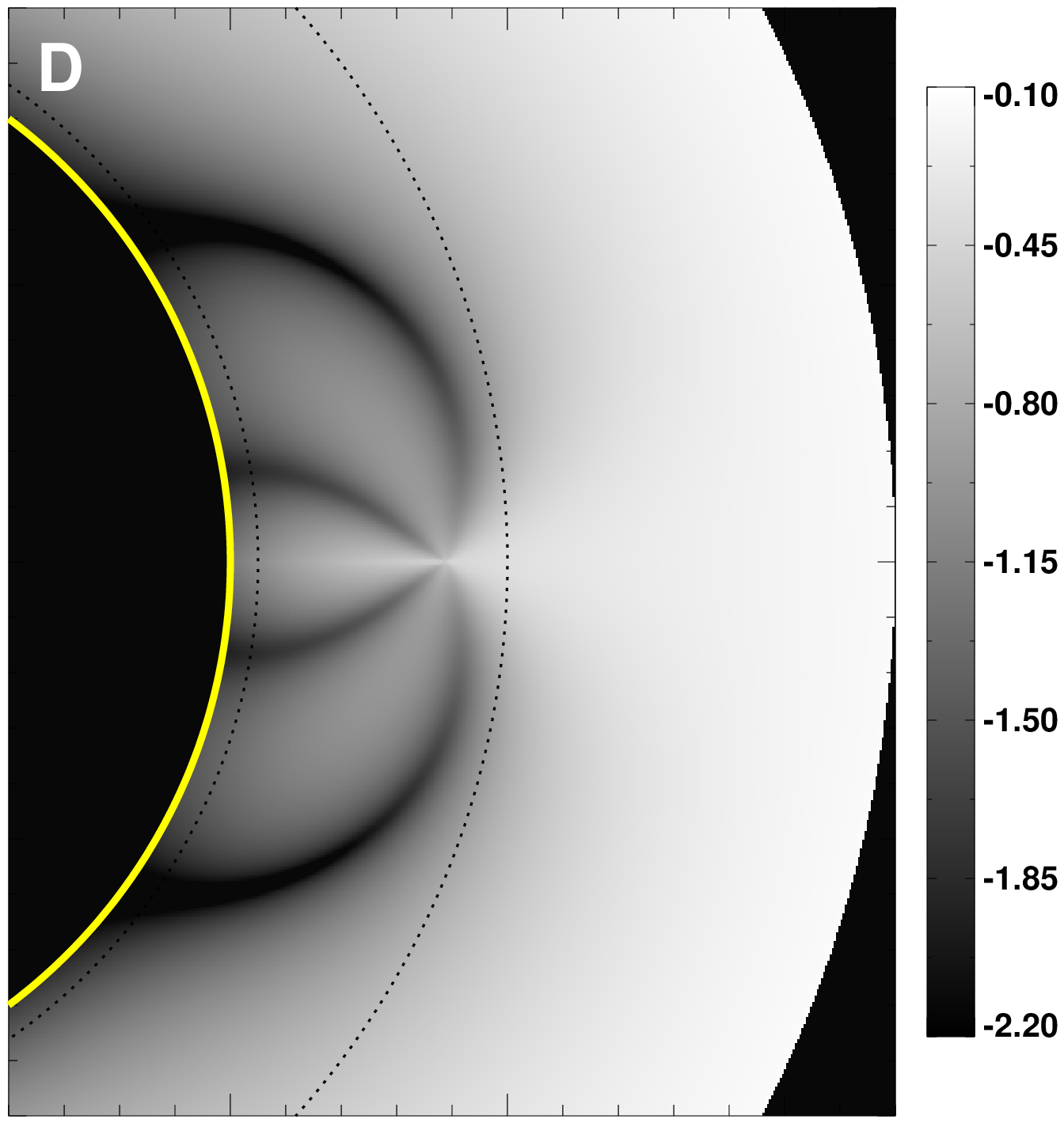}
              }

              \caption{2.5D axisymmetric models of a DS (A) and a PS
                (B) are shown with field line traces. Note that the
                field lines were chosen to illuminate topology and
                their density does not correspond to field
                strength. The nulls are marked with stars, and the
                separatrices as dashed or dash-dotted lines. The
                corresponding forward-modeled relative linear
                polarization, $L/I$, is shown below in (C), (D) in log
                scale with the field of view of the CoMP instrument,
                demarcated by the dotted
                lines. \label{fig:PS-SBS-mag-comp}}
\end{figure}

Double streamers and pseudostreamers are similar, but topologically
distinct, magnetic structures. Figure~\ref{fig:PS-SBS-mag-comp}A
and~\ref{fig:PS-SBS-mag-comp}B shows these two magnetic configurations
and their magnetic skeletons \citep{parnell2008}. Both models are
simple axisymmetric potential field source surface (PFSS)
extrapolations from a photospheric boundary magnetic field calculated
using spherical harmonics, and do not contain longitudinal field.

In the 2D cross section of the DS model shown in
Figure~\ref{fig:PS-SBS-mag-comp}A, there are two null points at the
upper source surface (marked by stars). Each null point forms the
upper tip of a cusp-shaped separatrix (dashed lines) that encloses
each closed loop volume below it. A trumpet-shaped volume of open
field is sandwiched between the two closed-field volumes, and is of
opposite polarity to the polar open field.

Unlike in a DS, in a PS there is no open field between the closed
field volumes that straddle the two PILs. Instead, in 2D, the
separatrices that bound the two closed field regions meet at a single
null. Outside of the closed field regions in the PS model, there are
two open field domains of like polarity. These open field domains meet
above the null point and are separated by the spine of the null. The
spine also extends downward from the null to separate the two closed
field regions from one another. In Figure~\ref{fig:PS-SBS-mag-comp}B,
the spines are shown with a dash-dot line and the separatrices are
dashed.

\begin{figure}\centering
  \epsscale{.80}
  \includegraphics[width=0.29\textwidth,clip=]{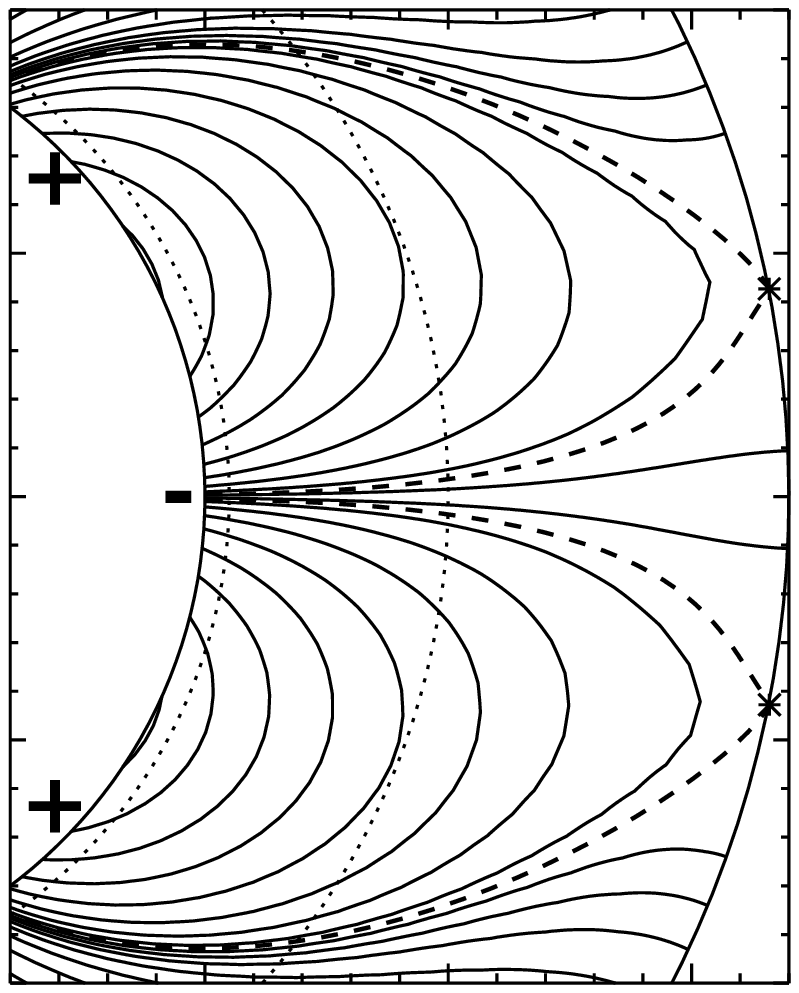}
  \hspace*{+0.01\textwidth}
  \includegraphics[width=0.2935\textwidth,clip=]{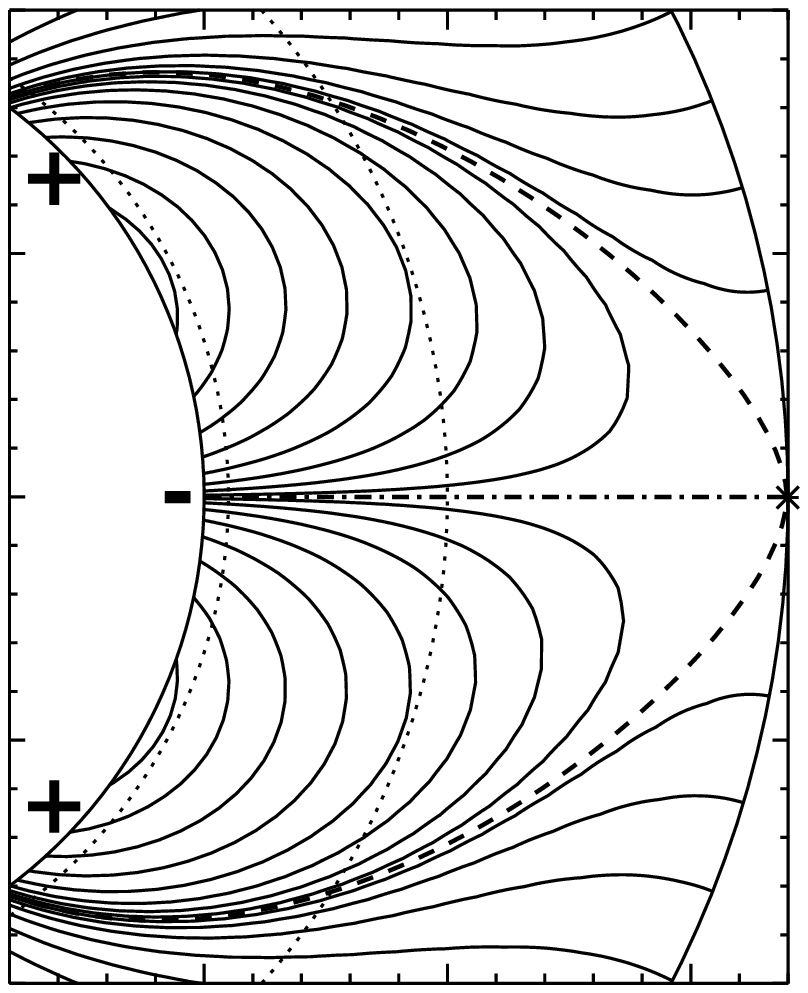}
  \hspace*{+0.01\textwidth}
  \includegraphics[width=0.29\textwidth,clip=]{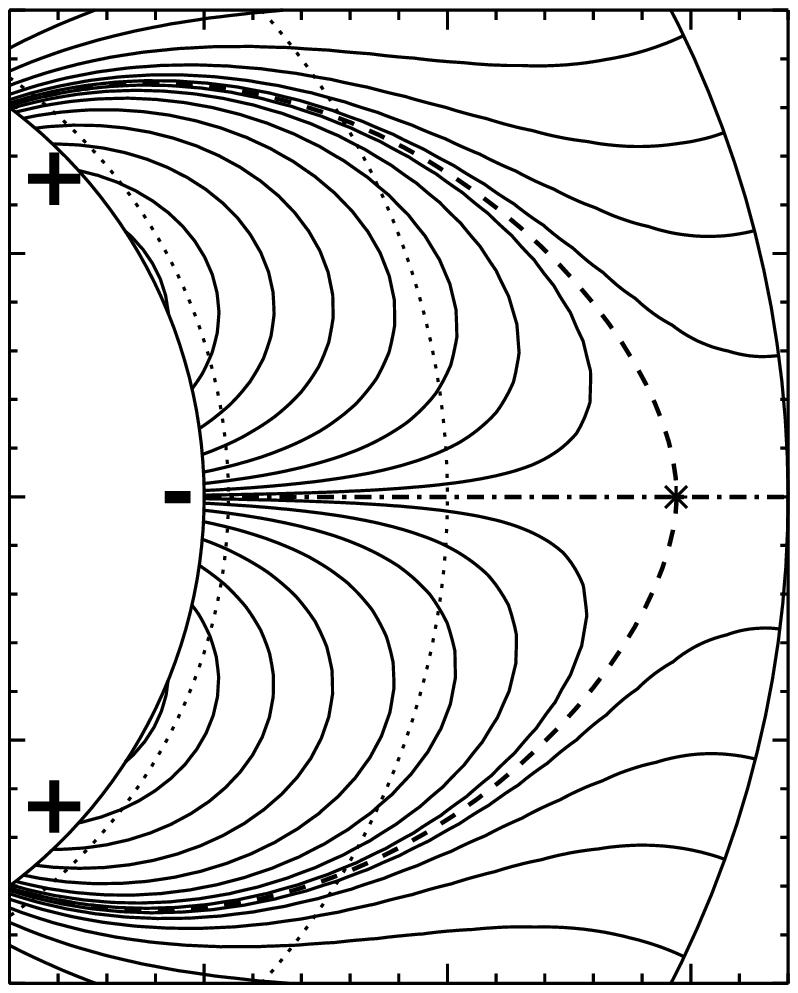}
  \caption{A cartoon showing the transition between DS
    (left) and PS (right) morphologies as longitudinal slices. Stars denote the
    location of the nulls. Field line density does not correspond to
    field strength. This figure is available as an animation in the
    electronic edition of the {\it Astrophysical
      Journal}.\label{fig:tot_skel}}
\end{figure}

A DS and a PS can be combined into a single hybrid structure with two
continuous PILs. The transition from a DS to a PS is accomplished by
the narrowing of the distance between the two separate DS null-lines
until they meet and merge into a single null. A decrease of magnetic
field strength of the central polarity between the two PILs on the
photosphere could cause this narrowing. Figure~\ref{fig:tot_skel}
qualitatively shows how this might occur in 3D by illustrating
longitudinal cuts in a highly simplified toy model of the system.

While our models capture the basic characteristics and skeleton of the
magnetic configurations, when present on the sun, such structures
would have a more complex topology. PILs on the sun are rarely
perfectly straight, currents are practically ubiquitous, and the
structure would be influenced by external features such as active
regions. \citet{Titov2012} present a detailed topological description
of several complex PSs which shows how these simple magnetic
structures quickly become quite complex in a more realistic
environment.

\subsection{Simulated coronal polarization}

We use the axisymmetric models shown in
Figure~\ref{fig:PS-SBS-mag-comp}A and \ref{fig:PS-SBS-mag-comp}B to
simulate coronal polarization, which is directly sensitive to the
magnetic field in the corona. Figures~\ref{fig:PS-SBS-mag-comp}C, and
\ref{fig:PS-SBS-mag-comp}D show the line-of-sight integrated simulated
relative linear polarization, $L/I=\sqrt{Q^2+U^2} / I$, where $I$,
$Q$, and $U$ are the Stokes parameters of the emission from the Fe
XIII transition at 10747 \mbox{\normalfont\AA}. This polarization is
primarily due to the Hanle effect in the saturated regime\footnote{For
  a more complete discussion of the Hanle effect in the saturated
  regime, see for example \citealt{casini1999},
  \citealt{Trujillo2001}, \citealt{casini2002}.}.

Beginning with the magnetic configurations shown in
Figures~\ref{fig:PS-SBS-mag-comp}A, and~\ref{fig:PS-SBS-mag-comp}B
respectively, we used the \textsf{FORWARD} suite of codes
\citep{judge2006, gibson2010, rachmeler2012, rachmeler2013} to
generate simulated polarization images using a simple spherically
symmetric hydrostatic temperature and density model
\citep{gibson1999}. The elongated dark features in these images mark
the locations of the Van Vleck inversions -- where the magnetic field
is at $\sim$$54^{\circ}$ from solar radial. A set of closed magnetic
loops in the plane of the sky results in two elongated Van Vleck inversion lines
where $L$ approaches zero\footnote{For more information on
  interpreting linear polarization observations see
  \citealt{rachmeler2012} and references therein.}.

In most observations of coronal cavities on the limb, where the PIL is
aligned with the line-of-sight, we see decreased $L/I$ above the PIL
due to field that is sheared or twisted into the line-of-sight
\citep{baksteslika2013}. Because the structures in our simple models
do not contain azimuthal field, this decreased $L/I$ is not
present. For our analysis, we focus on the locations of the Van Vleck
inversions, which are enough to distinguish between these two magnetic
configurations.

The differences between the $L/I$ signatures of these two magnetic
morphologies are clear. The DS (Figure~\ref{fig:PS-SBS-mag-comp}C)
shows two sets of roughly parallel Van Vleck inversions. Note that
although they converge at the null line on the source surface, in the
lower corona they are essentially parallel. This parallel nature is
consistent with other observations of coronal cavities that overlie
PILs when little or no line-of-sight field is
present. \citep{rachmeler2013,baksteslika2013}. In the PS, on the
other hand, the Van Vleck inversions clearly converge at the null
location. Although this specific convergence configuration is for
unsheared arcades, the addition of shear still results in Van Vleck
inversions that do not reach significantly higher than the
separator. Thus, these two structures are easily identifiable in $L/I$
observations provided that the separator is at low enough altitude to
appear in the field-of-view, and the PILs are along the line-of-sight.

\section{Observations} \label{sec:obs}

\begin{figure}

   \centerline{\hspace*{0.015\textwidth}
     \includegraphics[width=0.53\textwidth,clip=]{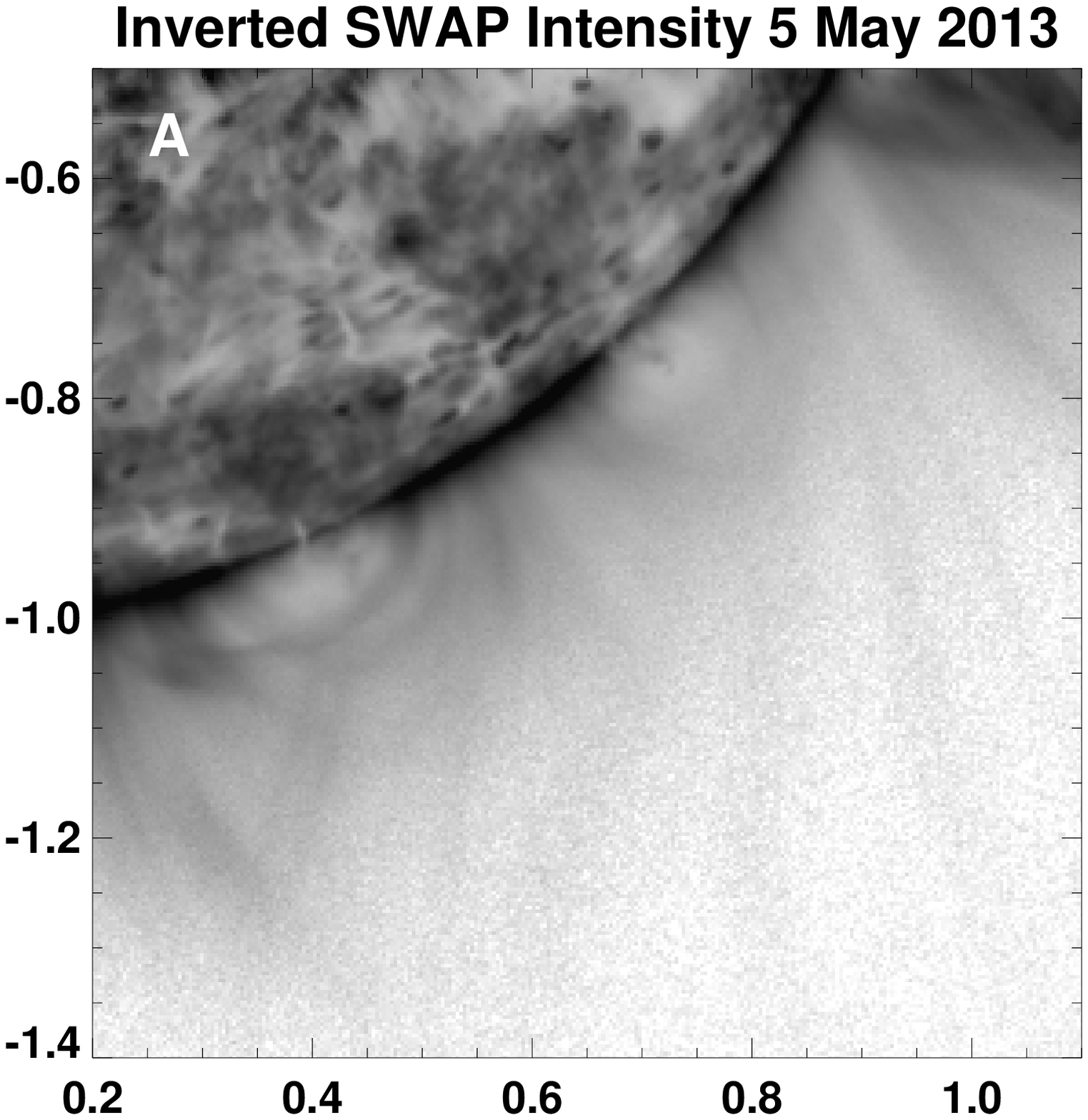}
               \hspace*{-0.05\textwidth}
               \includegraphics[width=0.53\textwidth,clip=]{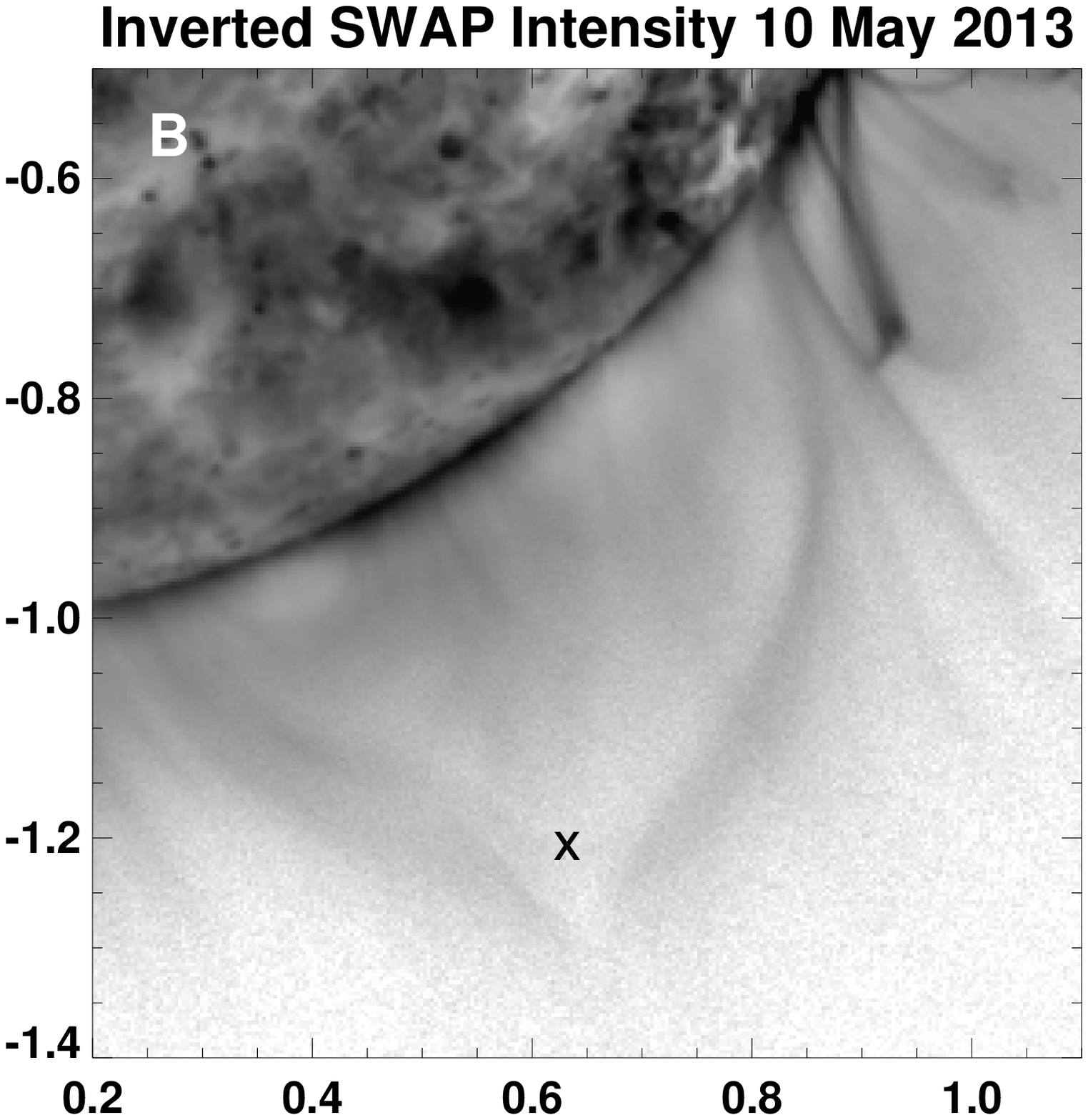}
              }
              \vspace{-0.05\textwidth}
   \centerline{\hspace*{0.015\textwidth}
     \includegraphics[width=0.53\textwidth,clip=]{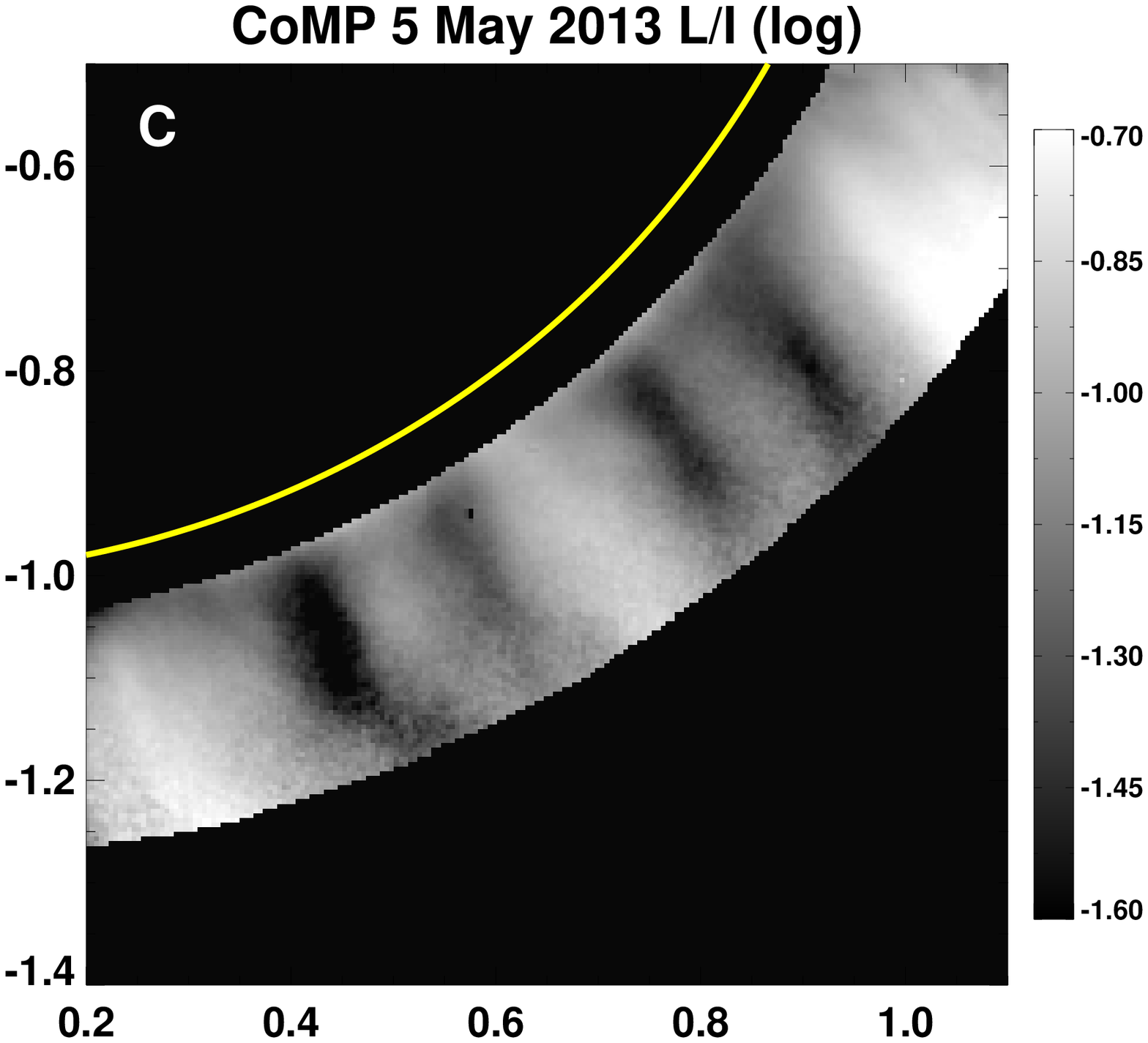}
               \hspace*{-0.05\textwidth}
               \includegraphics[width=0.53\textwidth,clip=]{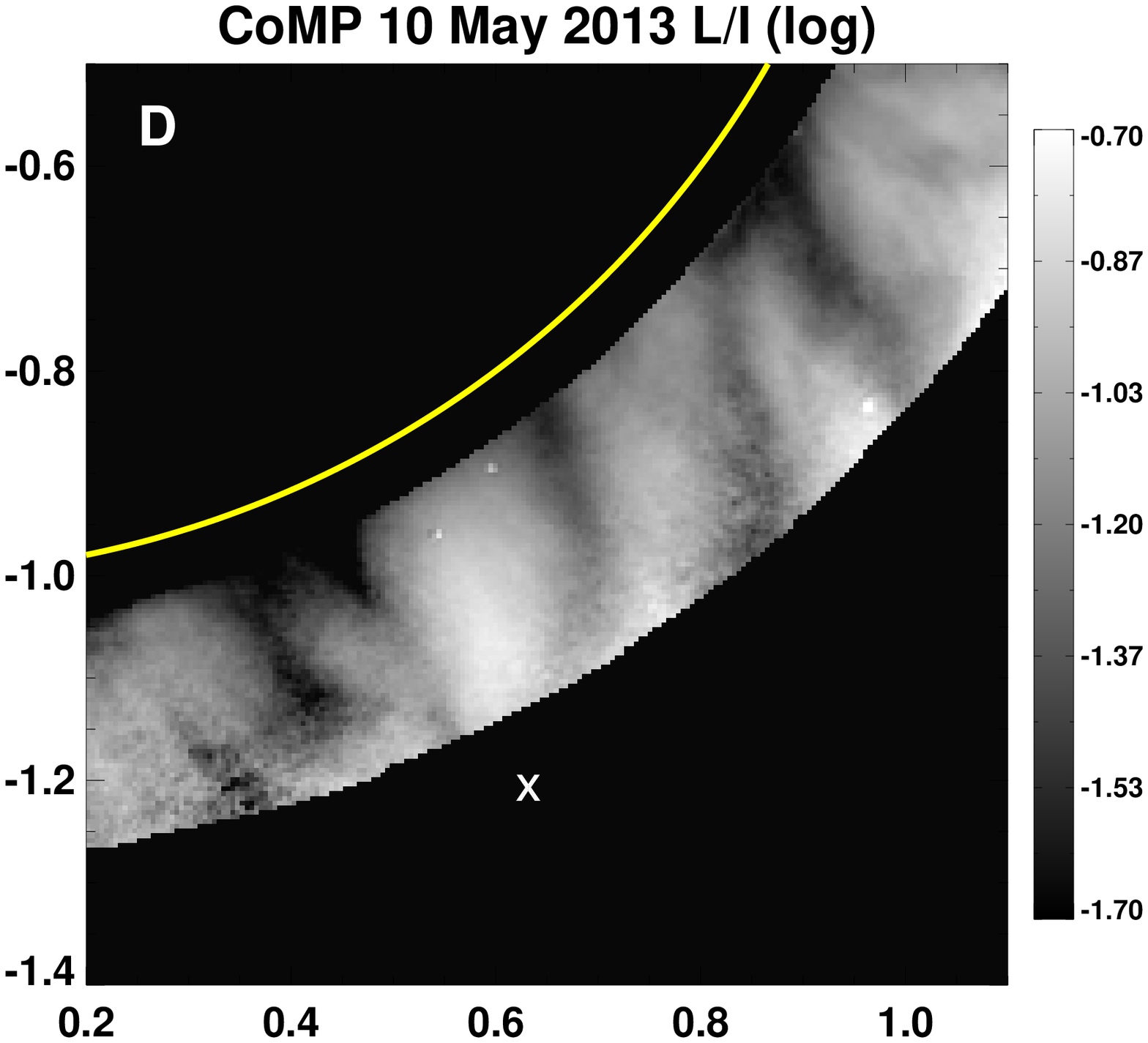}
              }

              \caption{Inverted SWAP (upper) and CoMP (lower)
                observations of the unified structure as a DS on
                5~May~2013 (left), and a PS on 10~May~2013
                (right). The `\textsf{x}' in B and D shows the
                approximate separator location for the PS. All length
                units are in $R_\odot$ from disk center. The SWAP
                images are available as an animation in the electronic
                edition of the {\it Astrophysical
                  Journal}. \label{fig:obs}}
\end{figure}

The observations presented here include data from the following
instruments: PROBA2/SWAP 174~\mbox{\normalfont\AA} (PRoject for
Onboard Autonomy 2/Sun Watcher using Active Pixel System detector and
Image Processing; \citealt{seaton2013, halain2013}); CoMP
10747~\mbox{\normalfont\AA} (Coronal Multichannel Polarimeter;
\citealt{tomczyk2008}) relative linear polarization ($L/I$); and
ChroTel (Chromospheric Telescope; \citealt{bethge2011}) H$\alpha$
6563~\mbox{\normalfont\AA}. Since CoMP is a ground-based instrument,
and only observes at specific times, we chose two CoMP observations to
mark the times the DS (19:00~UT on 5~May~2013) and PS (19:30~UT on
10~May~2013) were visible on the western limb of the sun. These CoMP
$L/I$ observations are shown in Figures~\ref{fig:obs}C and
\ref{fig:obs}D. Simultaneously obtained SWAP observations are shown in
Figures~\ref{fig:obs}A and \ref{fig:obs}B.

The intensity observations in SWAP appear to show a topological
transition from a DS to a PS. The DS structure is more ambiguous than
the PS structure. The 10~May observation clearly shows a cusp-shape
void (near the `\textsf{x}' in Figure~\ref{fig:obs}B) above the two
filament channels; bright structure is also clearly visible just
outside the void. The bright features indicate that two domains of
open field come together, and because this happens above two PIL's,
they are likely of the same polarity. Note that in
Figure~\ref{fig:obs}, the SWAP data is inverted, so bright emission is
dark. This cusp structure is likely a temperature effect suggesting
that the void contains plasma that is too hot to be seen by SWAP
\citep{seaton2013.2}. In the 5~May data, there is no clear indication
of open field between the cavities, but no cusp shape is visible.

The DS nature of the structure on 5~May is, however, clearly visible
in the CoMP data (Figure~\ref{fig:obs}C). The two sets of Van Vleck
inversions are distinct and roughly parallel. This is very similar to
the simulated emission from the analytical DS model
(Figure~\ref{fig:PS-SBS-mag-comp}C). It is interesting that these
streamers do not have decreased $L/I$ emission between the Van Vleck
inversion, indicating that any sheared or twisted field is either
entirely below the 1.05~$R_\odot$ CoMP occulter or not present
\citep{baksteslika2013}. The small size of the cavities in the
corresponding SWAP image indicate that the former is likely.

The PS observation in CoMP (Figure~\ref{fig:obs}D) is not as clear as
the DS observation. There are several Van Vleck inversion lines that
begin to converge like they do in Figure~\ref{fig:PS-SBS-mag-comp}D
but the convergence point (at or near the location of the
`\textsf{x}') is unfortunately located outside of the CoMP
field-of-view. There are six elongated dark structures visible in
Figure~\ref{fig:obs}D. The uppermost and lowermost are located outside
of the PS. The upper one is related to the edge-on loops seen in SWAP,
and the lower is a noise artefact. The inner four are the Van Vleck
inversions of the PS.  The quality of the data was also better on
5~May than 10~May due to weather conditions.

For both dates, it is the \emph{combination} of the SWAP and CoMP data
that points to the transition from a DS to a PS structure. On 5~May,
the CoMP data are clearly indicative of a DS, which is consistent with
the SWAP data although the SWAP data is ambiguous. On 10~May the
cusp-shape in the SWAP observation is a clear sign of two open field
domains coming together above two PILs, indicative of a PS
morphology. The CoMP data supports this with the Van Vleck inversions
being bent towards the location of the top of this cusp, although
there are still strong ambiguities in the CoMP observations, most
notably because the top of the PS lies above the CoMP field-of-view.

\begin{figure} \centering
    \includegraphics[width=\textwidth,clip=]{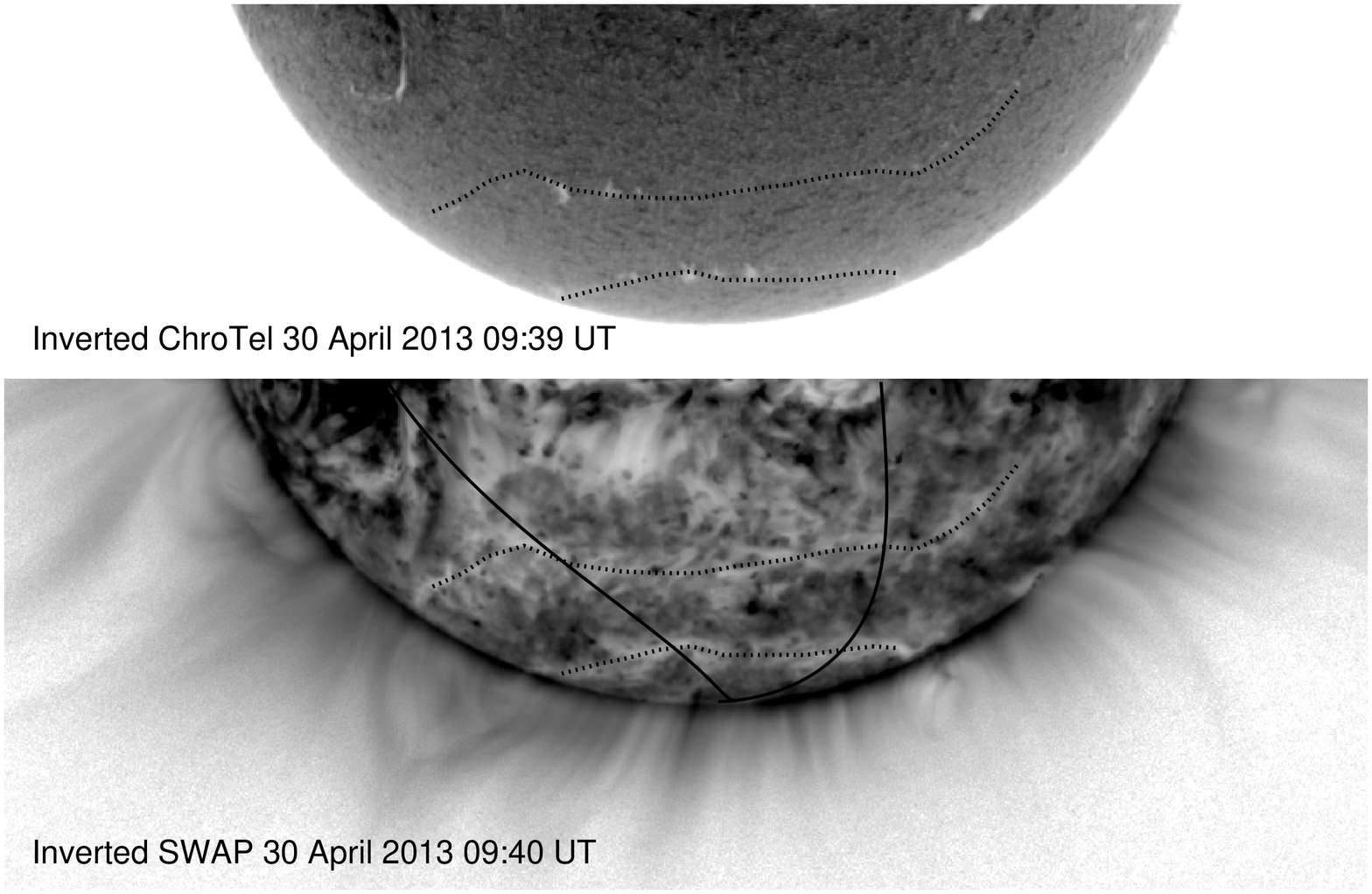}
    \caption{Inverted SWAP and ChroTel observations showing the two
      filament channels, which are traced with dotted lines. The solid
      lines show the de-rotated longitudes of the DS and PS as
      observed by CoMP on 5~May~2013 (right line) and 10~May~2012
      (left line). }\label{fig:filament}
\end{figure}

To determine the locations of the two filament channels, we used
medianed high signal-to-noise SWAP data obtained on 30~April~2013
around 09:40~UT, when the filament channels were located
on-disk. The longitudes at which the DS and PS were observed are
indicated by the solid lines on the SWAP image in
Figure~\ref{fig:filament}. The filament channels appear as two
elongated dark structures near the south pole (light in the inverted
SWAP image), indicated by the dotted lines in
Figure~\ref{fig:filament}. H$\alpha$ observations from ChroTel reveal
that neither channel is consistently filled with filament material.

The lack of clearly identifiable filament material makes it difficult
to identify the location of both filament channels precisely. We used
the standard PFSS extrapolation from 4~May~2013 at 12:04~UT from the
\textsf{SolarSoft} PFSS package to confirm the existence of two
continuous PILs. The extrapolation does indeed show two PILs with
closed arcade structures above them. 

The arcades increase in height from east to west along the channels,
but do not open into two separate streamers on the west end where our
observations show a DS. Indeed we would not necessarily expect to see
a DS as PFSS extrapolations are potential and do not accurately model
locations where electrical currents are important, such as near
filaments. The inclusion of such currents along a PIL (resulting in
sheared or twisted field) can cause an expansion of the overlying
arcade compared to PFSS extrapolations. The strength of the currents
can also vary along the PIL (Figure 3 by \citealt{yeates2012}). The
sheared field would increase the magnetic pressure, and thus the
outward-directed magnetic force, while the inward tension force from
the overlying arcade would remain the same. Thus, the equilibrium
magnetic configuration is inflated when shear or twist are introduced
to the system.

Furthermore, the open/closed field boundaries are strongly dependent
on the height of the source surface, the standard height being 2.5
R$_\odot$. However, depending on the solar activity, a source surface
as low as 1.5 R$_\odot$ may best fit the observations
\citep{Lee2011}. Lowering the source surface has the effect of
increasing the size of the open field domains and could also easily
result in open field between two large arcades. Thus, the lack of a DS
in the PFSS model does not preclude the presence of a DS in the true
corona.

Between May~5 and 10, there are two eruptions involving the lower
filament channel. The first eruption occurs at roughly
7~May~01:20-21:45~UT and the second at 9~May~02:45-13:45~UT. Both
occur primarily in the PS section of the structure. Neither disrupts
the underlying morphology of the hybrid system except temporarily
during the eruption itself.

\section{Discussion}

In this letter we have presented an observation and a simplified model
of a single hybrid magnetic structure containing both a side-by-side
double-streamer (DS) and a pseudostreamer (PS) along two continuous
filament channels. This morphological change is supported by a
combination of SWAP EUV images and CoMP linear polarization
measurements. While previous studies \citep{dove2011, baksteslika2013,
  rachmeler2013} have considered the coronal polarization signatures
of streamers, this letter presents the first research on coronal
polarization characteristics of PSs.

The CoMP data from 5~May~2013 (Figure~\ref{fig:obs}C) is consistent
with a DS structure, with two sets of roughy parallel Van Vleck
inversion lines, much like those shown in our analytic model
(Figure~\ref{fig:PS-SBS-mag-comp}C). Several days later, SWAP sees a
cusp-shape where two open field domains come together above two PILs,
which is a good indication of of a PS. The CoMP data from that time
shows two sets of Van Vleck inversion lines that begin to converge on
a location above upper boundary of CoMP (Figure~\ref{fig:obs}D). This
characteristic convergence towards the PS null or separator is
predicted by the \textsf{FORWARD} model for a PS
(Figure~\ref{fig:PS-SBS-mag-comp}D).

Despite the absence of continuous filament material over the full
extent of both channels (Figure~\ref{fig:filament}), PFSS
extrapolations confirm the presence of two continuous PILs. The
combination of observations and models presented here point to a
single hybrid structure containing both DS and PS magnetic
morphologies.

The solar wind from streamers is generally thought to be slow wind
\citep{gosling1981, strachan2002}, while there is some debate about
the nature of the wind from that originates in and around
pseudostreamers \citep{wang2007, riley2012, wang2012,
  panasenco2013}. As a result, as the hybrid structure rotates across the
solar disk, the transition from a DS to a PS may influence the
characteristics of the solar wind between the Sun and the Earth.

The open field at the center of the DS produces fast solar wind, while
the two streamers produce slow wind. The three streams may interact in
interesting ways once they reach the heliosphere, implying that the
solar wind from the DS alone could have complex structure.

Furthermore, reconnection in a helmet streamer occurs primarily in the
current sheet, while reconnection in the PS occurs primarily at the
separator. Because the reconnection in the PS occurs at a lower
height, and it is likely that the composition of the solar wind
originating from the two regions is different.

The change in magnetic configuration could also affect the stability
of the enclosed filament.  The lower height of the separator in the PS
structure likely results in a more rapid decrease in field strength
with height. This may reduce the height of the critical value of the
decay index, which determines the filament's vulnerability to eruption
via torus instability \citep{kliem2006, demoulin2010,
  zuccarello2012}. \citet{torok2011}, \citet{Titov2012}, and
\citet{lynch2013} have also shown that an eruption in one of the lobes
of a PS can easily trigger a sympathetic eruption in the other
lobe. Thus these hybrid structures may be more vulnerable to eruption
than streamers or double streamers alone.

Although this letter presents the first identification of this type of
hybrid structure, we do not believe they are an uncommon
phenomenon. Especially near solar maximum, when there are multiple
polar crown filaments \citep{mouradian1994, minarovjech1998} that are
slowly driven together due to the meridional flow, there is the
potential for similar structures to form. More work is needed to find
further instances of such structures, analyze their 3D topology in
detail, and investigate their heliospheric implications. 

\acknowledgments {D.B.S. and L.A.R. acknowledge support from the
  Belgian Federal Science Policy Office (BELSPO) through the
  ESA-PRODEX program, grant No. 4000103240.  S.J.P. acknowledges
  the financial support of the Isle of Man Government. The
  authors would like to thank Francesco Zuccarello, Sarah Gibson,
  Duncan Mackay, and Clare Parnell for helpful discussions. SWAP is a
  project of the Centre Spatial de Liege and the Royal Observatory of
  Belgium funded by BELSPO. CoMP data is provided courtesy of the
  Mauna Loa Solar Observatory, operated by the High Altitude
  Observatory (HAO), as part of the National Center for Atmospheric
  Research (NCAR). NCAR is supported by the National Science
  Foundation. ChroTel is operated by the Kiepenheuer-Institute for
  Solar Physics (KIS), at the Spanish Observatorio del Teide.}

{\it Facilities:} \facility{PROBA2}, \facility{STEREO},  \facility{SDO}





\end{document}